\documentclass[usenatbib,useAMS]{mn2e}

\usepackage[dvips]{graphicx}

\title[Crab Nebula Jet]{The Crab Nebula's Dynamical Age as Measured from its Northern Filamentary Jet}

\author[G. C. Rudie, R. A. Fesen, and T. Yamada]{G. C.
Rudie$^{1,2}$\thanks{E-mail: gwen@astro.caltech.edu (GCR);
fesen@snr.dartmouth.edu (RAF); yamada@astr.tohoku.ac.jp (TY)}, R. A. Fesen$^{2}$,
and T. Yamada$^{3}$\\ $^{1}$Department of Astrophysics, MC 105-24, Caltech, 
1200 E. California Blvd., Pasadena, CA 91125, USA\\
$^{2}$Department of Physics \& Astronomy, 6127 Wilder
Laboratory, Dartmouth College, Hanover, NH 03755, USA\\ 
$^{3}$Astronomical Institute, Tohoku University, Aoba-ku, Sendai 980-8578, JAPAN}

\begin{document}
\date{Accepted 2007 MONTH DAY. Received 2007 MONTH DAY; in original form 2007 July 30}
\pagerange{\pageref{firstpage}--\pageref{lastpage}} \pubyear{2007}

\maketitle

\label{firstpage}

\begin{abstract}

We present a deep [O~III] $\lambda\lambda$4959,5007 image of the northern
filamentary jet in the Crab Nebula taken with the 8.2m Subaru telescope.  Using
this image and an image taken with the KPNO 4m in 1988 \citep{fes93}, we have
computed proper motions for 35 locations in the jet. \textbf{The results suggest that when 
compared to the main body of the remnant, the jet experienced less outward acceleration 
from the central pulsar's rapidly expanding synchrotron nebula.} 
The jet's apparent expansion rate yields an undecelerated
explosion date for the Crab Nebula of 1055 $\pm$24 C.E., a date much closer to
the appearance of the historic 1054 C.E.\ guest star than the 1120 -- 1140 C.E.\
dates estimated in previous studies using filaments located within the
remnant's main nebula. \textbf{Our proper motion measurements suggest the jet likely formed during
the 1054 supernova explosion and represents the remnant's highest velocity
knots possibly associated with a suspected N-S bipolar outflow from the supernova
explosion.} 

\end{abstract}


\section{Introduction}

The first object in Messier's catalog of nebulous sources, the Crab Nebula
(M1), is a young galactic supernova remnant (SNR) and one of the most well
known astronomical objects. It possesses a variety of remarkable properties and
structures including: a highly complex structure of optically bright filaments
as seen in recent high-resolution Hubble Space Telescope images \citep{hes96,loll04},
an unusually luminous central pulsar (see Harden \& Seward 1984 and Rots et al.
2004 and references therein) with accompanying X-ray jets \citep{bri85}, and a
radio and optically bright synchrotron nebula which exhibits fine-scale
structural changes on timescales of just days or weeks \citep{sca69,sca70,hes95}. 

The Crab Nebula is generally considered the archetype of a pulsar driven
synchrotron wind nebula (PWN) and of the so-called  ``plerionic'' remnants in
general (see review by \citealt{dav85}). The spinning magnetic field of
the pulsar and strong relativistic wind of particles off the neutron star have
measurably accelerated the remnant's optical filaments such that their
velocities, when projected back, give a date of $\sim$1130 C.E. \citep{tri70,
nug98}.  This date is $\sim$80 years more recent than the 1054 C.E. date, the
year a ``guest star" associated with the Crab supernova is reported to have
been observed by the Chinese and other cultures in several ancient astronomical
texts \citep{cla77,ste02}. This apparent filament acceleration, confirmed by
several independent proper motion studies (see Table 1), due to an outward
pressure contained within the remnant's rapidly expanding PWN is described in detail by
\citet{TR70}.

%
%
\setcounter{table}{0}
\begin{table*}
\centering 
\begin{minipage}{140mm} 
\caption{Results from Previous Proper Motion Studies} 
\begin{tabular}{@{}lcccc@{}} 
\hline 
Reference &
Number of Knots & Image Epochs & Convergence Date & $\sigma$ (yr)\\
\hline 
Duncan
1939             & 20    & 1909, 1938                          & 1172        &
...    \\ 
Trimble 1968            & 132   & 1939, 1950, 1953, 1964, 1966, 1966
& 1140        & 15         \\ 
Nugent 1998             & 50    & 1939, 1960,
1976, 1992              & 1130        & 16         \\
\hline
\end{tabular} 
\end{minipage} 
\end{table*}
%
%

%
%
\setcounter{figure}{0}
\begin{figure*}
\centering
\includegraphics[width=.9\textwidth]{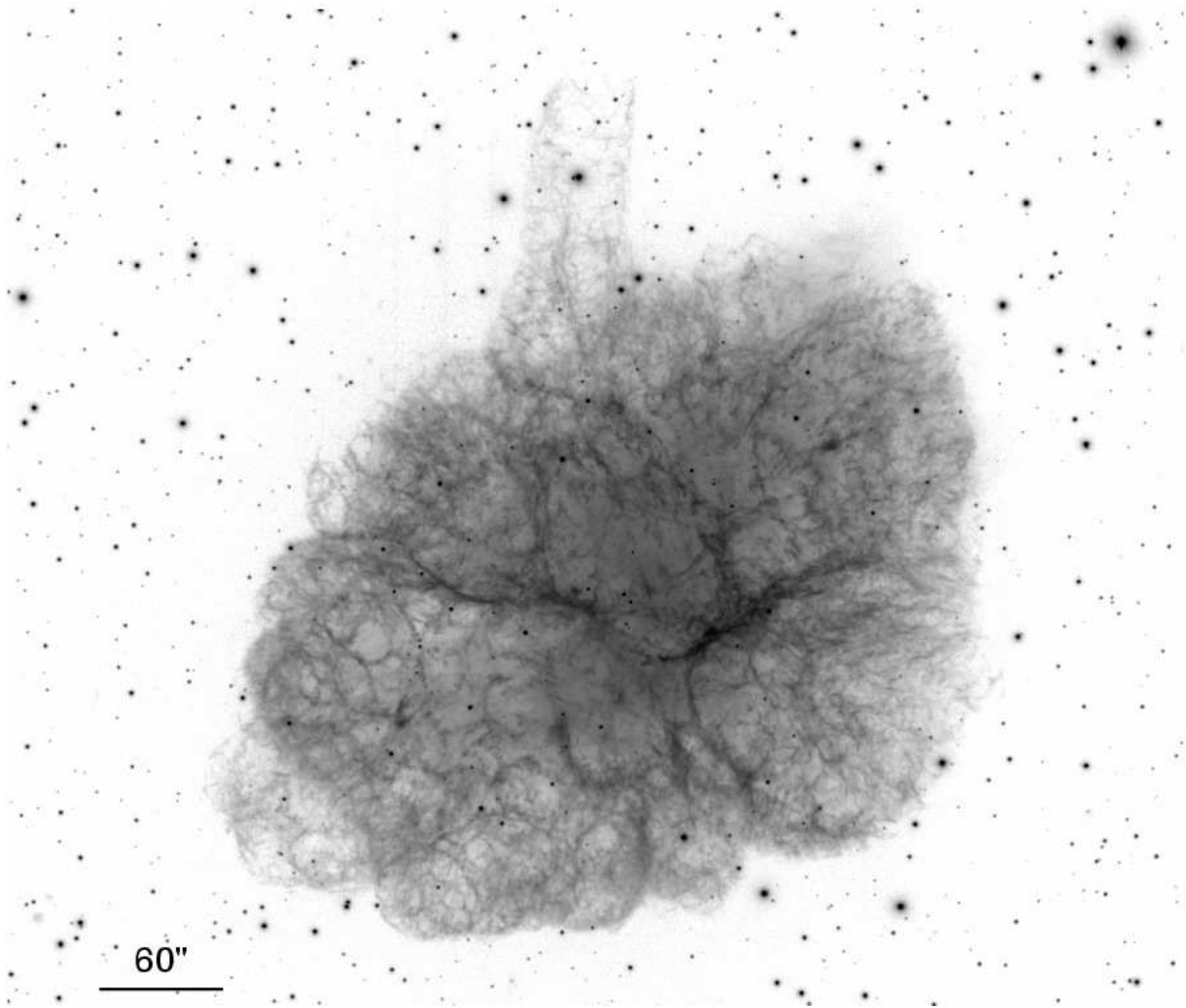}
 \caption{Subaru [O~III] image of the Crab Nebula using a log intensity stretch
to show the position of the northern filamentary jet in relation to the remnant's brighter 
inner filament complex.}
 \label{fig1}
\end{figure*}
%
%
%

A particularly intriguing feature of the Crab is a faint, northern filamentary `chimney'
or `jet' of emission filaments, which we will henceforth refer to simply as the
``jet'', but which should not to be confused with the physically smaller X-ray
synchrotron jets off the central pulsar.  First detected in the optical by
\citet{van70}, this optical filamentary jet consists of line-emitting
filaments, especially prominent in [O~III] $\lambda\lambda$4959,5007 \citep{che75}.
Although primarily a line-emission feature, very weak coincident non-thermal continuum
emission was subsequently discovered, first in the radio \citep{vel84} and then
in the optical \citep{wol87}, suggesting the jet itself is also filled with
relativistic particles off the pulsar. 

One especially puzzling aspect of the jet is its morphology. The first deep
[O~III] images of the jet showed it to be a complex network of filaments that
appear to be cylindrically symmetric, but with a central axis
unaligned with either the Crab's center of expansion or pulsar \citep{gul82}.
Follow-up imaging and kinematic studies \citep{shu84,ver85,mar90} lead to several proposed 
explanations including a mass loss trail from a red giant progenitor
\citep{bla83}, interaction of ejecta with an interstellar
cloud \citep{mor85}, and transverse synchrotron driven shocks \citep{san97}.  However,
preliminary proper motion studies of the jet indicated ordinary
radial motions away from the center of expansion \citep{fes86, fes93}.

Due to the jet's greater distance from the pulsar and the remnant's 
synchrotron nebula, the jet may have
experienced much less acceleration than most other regions of the remnant.
In that case, the jet might lead to a dynamically greater remnant age leading to
a date closer to the 1054 C.E. historic supernova sighting.

Here we present a deep and high resolution [O~III] image of the Crab jet taken
with the 8.2 m Subaru telescope. This image shows a
network of extremely fine and detailed filaments. We have compared this image
to a 1988 epoch image obtained with the KPNO 4m telescope \citep{fes93} to
obtain improved proper motion measurements of its optical filaments and an estimate of the
jet's expansion age. 


\section{Observations and Analysis}

Optical images of the Crab Nebula were
obtained 10 October 2005 using the 8.2 m Subaru telescope located on Mauna
Kea.  The image was taken using the Suprime-Cam Prime Focus Camera \citep{miy02} 
with the `NB497' narrow-band filter \citep{hay04} centered at approximately 4977 
\AA \ (FWHM of 77 \AA). A series of 180 s exposures taken under good seeing 
conditions (FWHM = 0.8$''$) were assembled to form a mosaic image of the entire 
Crab Nebula with a combined exposure time of 900 s. Figure~1 shows the full mosaic 
of the Crab with Figure~2 showing an enlarged section of the image centered on the 
northern jet feature.

%
%
\setcounter{figure}{1}
\begin{figure}
\includegraphics[width=0.5\textwidth]{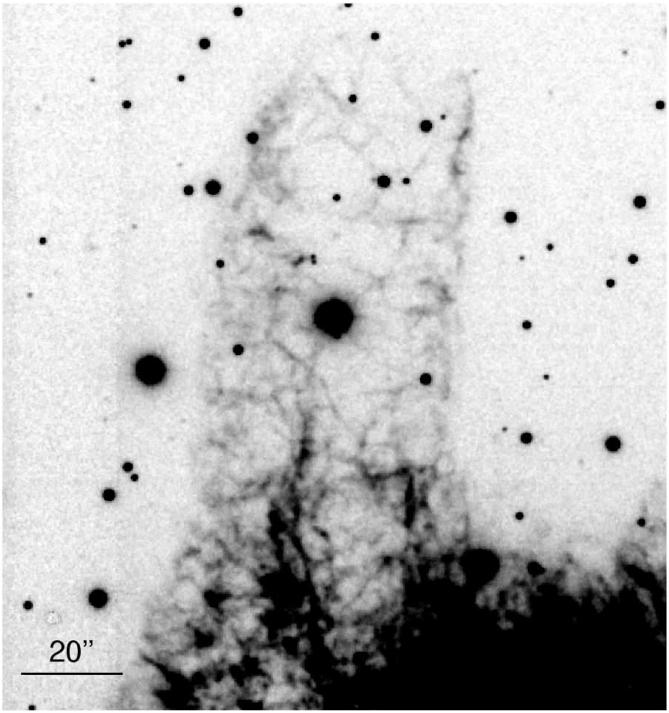}
 \caption{An enlarged section of the 2005 Subaru [O~III] image of the Crab 
Nebula showing its northern jet.}
 \label{fig2}
\end{figure}
%
%
%

We compared this Subaru 2005 image to a similar [O~III] image taken on 10 November 1988 
obtained using the KPNO Mayall 4 m telescope. The 1988 image is the same one previously used by
\citet{fes93} to estimate the proper motions of the jet. Here we chose 
35 emission knots within the upper portion of the jet (see Fig.\ 3) well detected on both
images for proper motion measurement. The 2005 Subaru data were reduced using the 
Suprime-Cam Deep field Reduction package which includes bias subtraction,
flat-fielding, distortion and atmospheric dispersion correction, and sky subtraction
\citep{yag02,ouc04}. The 1988 image was reduced using standard IRAF\footnote{
The Image Analysis and Reduction Facility is distributed by the National Optical 
Astronomy Observatories, which are operated by the Association of Universities 
for Research in Astronomy, Inc., under cooperative agreement with the National 
Science Foundation.} image reduction routines including bias subtraction, 
flat-fielding, and cosmic-ray rejection.

%
%
\setcounter{figure}{2}
\begin{figure}
\centering
\includegraphics[width=0.45\textwidth]{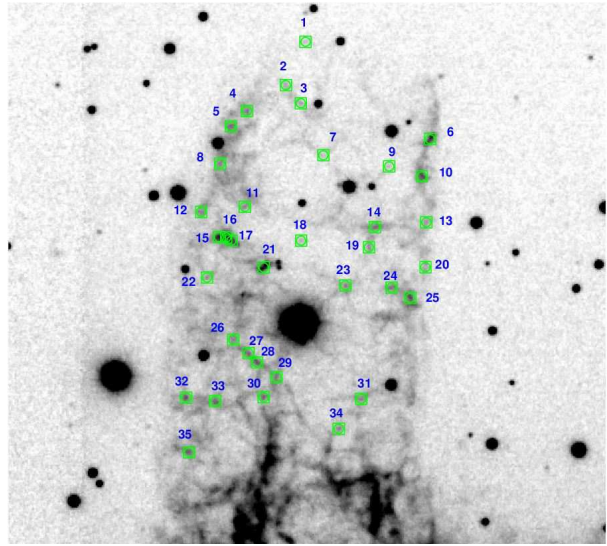}
 \caption{Subaru [O~III] image of the northern jet 
     with the 2005 positions of the 35 measured jet ejecta knots marked.}
 \label{fig3}
\end{figure}
%
%
%

World Coordinate System (WCS) information was applied to the images  in a two
step process.  First, stellar positions from the full 2005 image were compared
with the USNO B1.0 Star Catalog yielding a total of 232 matched fiducial stars.
The RMS deviation in the fit was 0.2$''$ in both RA and Dec. 

The WCS fit for the 1988 image was determined in a slightly different manner.
The 1988 jet image did not include the main body of the Crab and is slightly less
deep than the 2005 image. Consequently, the total possible number of USNO
fiducial stars in the 1988 imaged field is considerably fewer than on the
Subaru image. For this reason, we correlated the 2005 imaged stellar field with
the smaller 1988 image. The IRAF task `starfind' was used to determine the WCS
positions of stars in the jet region in the 2005 image. Then, from these
positions, eight stars were selected due to their low proper motions and PSFs.
These stars then were used as the fiducial ties between the 2005 WCS data and
the 1988 WCS image.  The resulting RMS deviation between the WCS fits of the
2005 and 1988 images was 0.1$''$ in RA and Dec and we believe the total
error in the WCS matching of the two images to be less than 0.2$''$.

Both images were projected onto tangent planes, with proper
motions computed from the corrected WCS coordinates.
Knot and filament positions in the jet were visually centroided on the 1988 and
2005 images.  Features were selected based mainly on the ease with which they 
could be reliably centroided and because they exhibited negligible shape and 
brightness variations apparent over the 17.1 year time interval between the 
two images.

The absolute accuracy of our proper motion measurements was limited by the
lower resolution of the 1988 image (FWHM = 1.0$''$) and by the
often diffuse nature of the knots and filaments.  We estimate the errors of our
centroided jet knot positions to be generally better than 0.3$''$ on the
1988 image and 0.2$''$ on the 2005 image.  The resulting proper motion
estimates should be more reliable and accurate than the prior study of
\citet{fes93} of the jet principally due to the deeper and higher resolution of
the 2005 jet image and a longer time interval (17 yr) between images
\citep{fes86, fes93}.

\section{Results}


Our primary goal was to investigate whether the jet's knots and filaments
experienced less acceleration due to their greater distance from the Crab
pulsar and its rapidly expanding synchrotron nebula.  In order to test and
verify the accuracy of our measurement technique for the jet knots and
filaments, we also included measurements for several emission knots in the
bright, inner portion of the remnant studied by \citet{tri68} and
\citet{nug98}. For these knots, we made measurements on the 2005 Subaru image
and also an image taken 28 November 1959 by N.\ Mayall with the 120-inch 
Lick Telescope.

Proper motion measurements were computed for a total of 140 knot and filament
positions distributed throughout the main nebula and the northern jet.
Measurements for knots located within the main body of the remnant resulted in
proper motion estimates ranging between 0.04$''$ and 0.21$''$ yr$^{-1}$
corresponding to transverse velocities of $350 - 1820$ km s$^{-1}$ at a
distance of 1.83 kpc \citep{dav85}. Looking only at those knots measured in
previous studies, projected back motions yielded estimated explosion dates 
of 1136 C.E.\ for 23 of Nugent's knots and
1146 C.E.\ for 42 of Trimble's knots, consistent with their reported
convergence dates of 1130 C.E. and 1140 C.E., respectively \citep{tri68,
nug98}. 

Measurements of 35 knots located in the northern half of the jet (see Fig.\ 3)
give a range of proper motion estimates of $0.19'' - 0.29''$ yr$^{-1}$,
corresponding to a transverse velocity range of 1650 -- 2520 km s$^{-1}$ at a
distance of 1.83 kpc \citep{dav85}. We also examined the motion of each
individual knot and projecting their positions back into the nebula.  For each
of these knots, we obtained an estimated convergence date back to the remnant's
center.  We chose the remnant's center of expansion as estimated by
\citet{nug98} and derived the convergence date as the year when the knot's backward
motion was closest to this expansion point.  The resulting convergence dates
are listed in Table 2. The average of these dates suggests an explosion date of
1050 C.E.\ with a standard deviation of 27.5 yr. 

Differences in the positional measurement uncertainties for the 35 jet knots
selected lead to several knots missing the expansion point by larger values
than the average of the sample. Therefore, in order to try to improve our
convergence date, we      selected a smaller knot sample which included just
those knots which showed the most radial proper motions, i.e.,  motions which
when projected back fell within 10$''$ of the \citet{nug98} center of
expansion.  Out of the 35 knots measured in the top of the jet, 22 had proper
motions which projected within this radius. This knot sample gave an average
convergence date of 1055 C.E.\ with a standard deviation of 24 yr. Figure 4
shows the scatter in the date given by all 34 measured jet knots with the
larger symbol (dots) indicating those which projected closest to the expansion
center.

%
%
\setcounter{table}{1}
\begin{table*}
\centering
 \begin{minipage}{115mm}
 \caption{Proper Motion Measurements for Jet Knots}
 \label{tab2}
 \begin{tabular}{@{}lcrrcccc@{}}
\hline
Knot& Convergence\footnote{Estimated date of knot's closest
(least squared distance) to the center of expansion}& D$_{CE}$
\footnote{The projected shortest distance between the knot and the center of expansion. 
} & $\mu_{x}~~~~$ & $\mu_{y}$ 
& $\mu_{T}$ & $\alpha$ (J2000)\footnote{Knot J2000 coordinates on the 2005 Subaru image.} & $\delta$ (J2000)$^{c}$
\\
ID&Date&($''$)~ &  ($''$ yr$^{-1}$) &  ($''$ yr$^{-1}$)& ($''$  yr$^{-1}$)&
(h~ m~ s)&($^{\circ}$ ~ $'$ ~ $''$) \\
\hline
   1  &  1095 &   7.276  &  0.02~~~ &   0.29  & 0.29 &  05 34 33.692 &+22 05 13.98   \\
   2  &  1066 &   4.404  &  0.01~~~ &   0.27  & 0.28 &  05 34 33.971 &+22 05 05.45   \\
   3  &  1092 &  11.639  &  0.03~~~ &   0.28  & 0.28 &  05 34 33.759 &+22 05 01.88   \\
   4  &  1065 &   1.344  &  0.03~~~ &   0.27  & 0.27 &  05 34 34.523 &+22 05 00.36   \\
   5  &  1090 &  12.195  &  0.04~~~ &   0.27  & 0.27 &  05 34 34.749 &+22 04 57.36   \\
   6  &  1054 &   9.951  & -0.02~~~ &   0.26  & 0.26 &  05 34 31.929 &+22 04 54.81   \\
   7  &  1059 &   3.076  &  0.01~~~ &   0.26  & 0.26 &  05 34 33.440 &+22 04 51.64   \\
   8  &  1015 &  15.081  &  0.05~~~ &   0.24  & 0.25 &  05 34 34.901 &+22 04 49.99   \\
   9  &  1035 &   2.589  & -0.01~~~ &   0.25  & 0.25 &  05 34 32.508 &+22 04 49.44   \\
  10  &  1044 &   6.287  & -0.02~~~ &   0.25  & 0.25 &  05 34 32.045 &+22 04 47.47   \\
  11  &  1048 &   7.347  &  0.03~~~ &   0.24  & 0.25 &  05 34 34.552 &+22 04 41.48   \\
  12  &  1067 &   4.962  &  0.04~~~ &   0.25  & 0.25 &  05 34 35.171 &+22 04 40.49   \\
  13  &  1039 &  14.228  & -0.03~~~ &   0.24  & 0.24 &  05 34 31.983 &+22 04 38.45   \\
  14  &  1050 &   7.636  & -0.01~~~ &   0.24  & 0.24 &  05 34 32.709 &+22 04 37.49   \\
  15  &  1023 &   9.589  &  0.04~~~ &   0.23  & 0.23 &  05 34 34.918 &+22 04 35.54   \\
  16  &  1001 &  10.281  &  0.04~~~ &   0.22  & 0.23 &  05 34 34.816 &+22 04 35.45   \\
  17  &  1029 &  11.021  &  0.04~~~ &   0.23  & 0.23 &  05 34 34.740 &+22 04 34.83   \\
  18  &  1028 &   9.539  &  0.02~~~ &   0.23  & 0.23 &  05 34 33.755 &+22 04 34.83   \\
  19  &  1004 &  11.673  & -0.01~~~ &   0.22  & 0.23 &  05 34 32.796 &+22 04 33.51   \\
  20  &  1038 &  13.634  & -0.02~~~ &   0.23  & 0.23 &  05 34 31.995 &+22 04 29.60   \\
  21  &  1083 &   6.839  &  0.03~~~ &   0.24  & 0.24 &  05 34 34.285 &+22 04 29.55   \\
  22  &  1085 &  10.776  &  0.05~~~ &   0.24  & 0.24 &  05 34 35.086 &+22 04 27.58   \\
  23  &  1087 &   4.165  &  0.00~~~ &   0.24  & 0.24 &  05 34 33.127 &+22 04 26.01   \\
  24  &  1011 &   3.100  & -0.01~~~ &   0.22  & 0.22 &  05 34 32.477 &+22 04 25.57   \\
  25  &  1020 &  18.716  & -0.03~~~ &   0.22  & 0.22 &  05 34 32.209 &+22 04 23.55   \\
  26  &  1081 &   5.428  &  0.04~~~ &   0.22  & 0.23 &  05 34 34.716 &+22 04 15.33   \\
  27  &  1024 &  12.762  &  0.01~~~ &   0.21  & 0.21 &  05 34 34.498 &+22 04 12.67   \\
  28  &  1028 &  11.971  &  0.01~~~ &   0.21  & 0.21 &  05 34 34.381 &+22 04 10.86   \\
  29  &  1051 &   8.362  &  0.01~~~ &   0.21  & 0.21 &  05 34 34.103 &+22 04 07.95   \\
  30  &  1055 &   1.637  &  0.02~~~ &   0.21  & 0.21 &  05 34 34.284 &+22 04 04.00   \\
  31  &  1076 &   6.288  &  0.05~~~ &   0.21  & 0.21 &  05 34 35.381 &+22 04 03.93   \\
  32  &  1063 &   4.861  & -0.00~~~ &   0.21  & 0.21 &  05 34 32.904 &+22 04 03.66   \\
  33  &  1067 &   7.809  &  0.04~~~ &   0.21  & 0.21 &  05 34 34.972 &+22 04 03.23   \\
  34  &  1007 &   0.479  &  0.01~~~ &   0.19  & 0.19 &  05 34 33.222 &+22 03 57.80   \\
  35  &  1066 &  11.303  &  0.05~~~ &   0.19  & 0.20 &  05 34 35.343 &+22 03 53.13   \\ 
\hline
 \end{tabular}
 \end{minipage}
\end{table*}
%
%

%
%
\setcounter{figure}{3}
\begin{figure}
\centering
\includegraphics[width=0.35\textwidth]{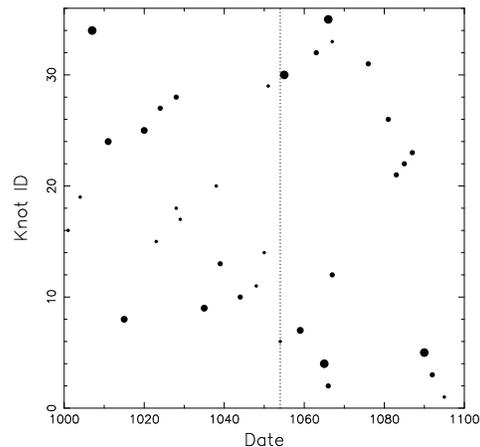}
 \caption{Plot of the computed convergence date of each knot upon the center of expansion. The size of the dot 
	in the figure corresponds inversely to the distance between the traced back motion of the knot and the center of expansion.
	Thus the largest dots correspond to those knots which project back closest to the center of expansion.}
 \label{fig4}
\end{figure}
%
%

In order to help judge the meaning of these northern jet proper motion results, proper motions
estimates were also made from the same images for knots in the southernmost part of
the jet and outside of the jet to the east and west, along the northernmost periphery of
the Crab's optical structure.  The results of these measurements are listed
in Table 3. These knot measurements yielded convergence dates
considerably earlier than those obtained from knots located in the main part of
the remnant but still later than the date we derived from the northernmost jet
knots.  Specifically, the 17 measured knots located in the base of the jet
whose projected-back proper motions fall within 10 arcseconds of the center of
expansion give a convergence date of 1081 $\pm$37 yr. The 22 measured knots
located to the east and west of the jet's base whose projected-back proper
motions also fall within 10 arc seconds of the center of expansion give a
convergence date of 1077 $\pm$45 yr. 

%
\setcounter{table}{2} \begin{table*} \centering \begin{minipage}{115mm}
\caption{Summary of Proper Motion Measurements} \label{tab3}
\begin{tabular}{@{}lcccccc@{}} \hline Region & r range &Number &  Image Epochs
&  $\mu_{T}$ & Convergence & $\sigma$\\ & ($''$)  & of Knots\footnote{ The
number of knots quoted here reflect the number of knots whose projected-back
proper motions fell within 10$''$ of the center of expansion. 
}  &   &  ($''$ yr$^{-1}$) & Date & (yr) \\ \hline Top of Jet
&  190 -- 270    &  22 & 1988.7 -- 2005.8   & 0.19 -- 0.29  & 1055  & 24    \\
Base of Jet           & 120 -- 160    &  17    & 1988.7 -- 2005.8   & 0.14 --
0.17  & 1081  & 37    \\ SE \& SW of Jet Base  &  110 -- 180    &  22    &
1988.7 -- 2005.8   & 0.11 -- 0.19  & 1077  & 45    \\ Whole SNR, Nugent     &
60 -- 180    &  23    & 1960.7 -- 2005.8   & 0.06 -- 0.20  & 1136  & 68    \\
Whole SNR, Trimble\footnote{Trimble's filament proper motions computed adopting
Nugent's center of expansion.}    &   40 -- 180    &  42    & 1960.7 -- 2005.8
& 0.04 -- 0.21  & 1146  & 54    \\ \hline \end{tabular} \end{minipage}
\end{table*}
%
%

In Figure 5, we show the projected motions of all 35 northern jet knots for an
expansion date of 1055 C.E., both for our measured knot proper motion
paths (right hand figure), and then assuming the motions of the knots were
completely radial, tracing perfectly away from the center of expansion as
defined by \citet{nug98} (left hand figure).  Overall, the observed motions of
knots within the jet region suggest a nearly radial expansion and trace back to
a location very close to the remnant's nominal center of expansion. Our
estimated positional measurement errors could have resulted in up to a
3$^{\circ}$ change in direction which would account for a substantial fraction
of the apparent non-radial motions seen in the right hand figure.  Thus, despite
the jet's strongly overall non-radial appearance with its long axis mis-aligned
with the remnant's center, individual jet knots and filaments appear to move in a radial fashion
away from the remnant's estimated explosion center. 

Two additional things should be noted from Figure 5. The first is that
a substantial number of jet knots appear to cross at a position north of
the center of expansion. Secondly,
knots located along the jet's western limb tend to show a convergence point
east of the center of expansion, while knots along the jet's eastern side appear
to favor a  convergence point west of the center of expansion.  Both these effects might
indicate a small, non-radial component to the jet's expansion which will be
addressed below.

%
%
\setcounter{figure}{4}
\begin{figure}
\centering
\includegraphics[width=0.45\textwidth]{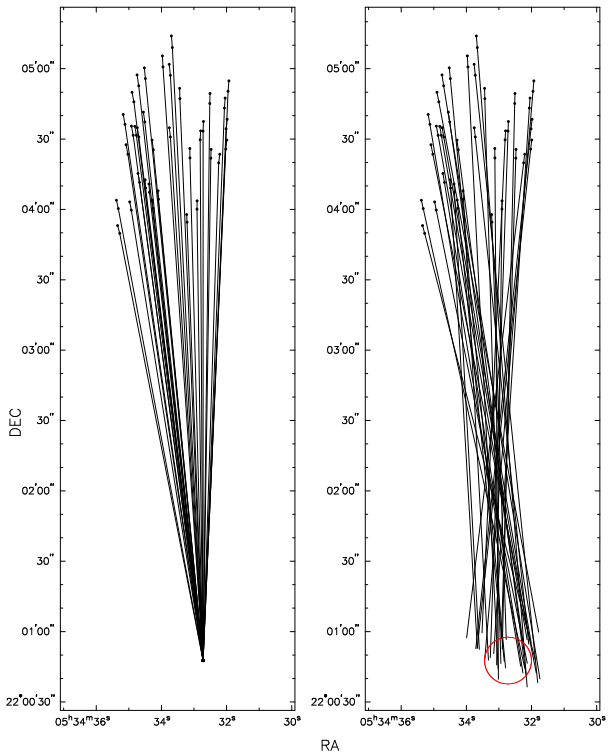}
 \caption{Plots of the extrapolated proper
motions of the jet knots. Both plots show two points, the measured positions
of the knots in 2005 and 1988. The lines in the left hand plot shows what the
motions would look like if they were perfectly radial. The lines in the right
hand plot show the true measured motions of the knots. The red circle is 10$''$ 
in radius centered on the center of expansion.}
 \label{fig5}
\end{figure}
%
%

\section{Discussion}
 
\subsection{Dynamical Age of the Crab Nebula}

Our computed convergence date for the jet is 1055 $\pm 24$ C.E., in excellent
agreement with historical records which indicate the Crab SN occurred in the
year 1054.  We attribute the reason for the difference between our result and
earlier age estimates to be chiefly the jet's northernmost knots experiencing
little appreciable acceleration by the pulsar's rapidly expanding
synchrotron nebula. All previous age estimates based upon proper motion studies
used knots and filaments located within the main body of the Crab remnant.
Consequently, interior knot motions reflected both an expansion due to the SN
event itself and a post-SN acceleration resulting from the interaction of the pulsar
wind nebula.  Although the 1054 explosion date for the Crab Nebula's SN has not
been in serious doubt, it is nonetheless valuable to have a direct
confirmation based on expansion dynamics.

Although the jet's knots and filaments appear not to have been significantly
accelerated radially away from the remnant's central region, our measured jet
knot proper motions may indicate a small, non-radial component.  As shown in
Figure 5 (right hand panel), the projected motions of the knots intersect north
of the position of the center of expansion. This effect may not represent the actual 
nature of the jet as there are several sources of error which could contribute to this motion.
First, it is possible that there may be a slight rotational shift between the
two World Coordinate System fits for the images.  Because the original motions
of the jet ejecta were so close to directly northward, a slight shift east or
west would cause a much more noticeable change in the E-W component of their
proper motion than would a similarly sized shift in the N-S direction. Thus, the
directions of the knots would be greatly affected by such a shift, while the
velocities would show considerably less error.  Secondly, an error in
centroiding the knots can account for a slight shift as well.  Our computed
errors show a change in direction of approximately 3$^{\circ}$ is possible
which could account for the non-radial motions of the knots.  

On the other hand, it is possible that the measured motions do reflect the  
jet's true expansion properties. The jet is essentially a hollow structure and as shown by
\citet{vel84} and \citet{wol87}, the jet does contain some faint synchrotron
emission, most visible at its southern base.  If this plasma induced an
acceleration that was largely against the jet's filamentary walls, this might act to
expand the jet in the east-west plane and explain why the
northwestern jet knots end up east of the center of expansion while most knots
located at the northeastern portion of the jet project back west of the center
of expansion.  Since our derived expansion date for the jet is most sensitive
to north-south motions, such an acceleration would not significantly effect our
estimated dynamical age. 

\subsection{The Nature of the Jet} 

Our proper motion measurements provide a valuable constraint on the 
origin of the Crab's northern jet of ejecta. A successful jet formation model should
explain the jet's non-radial appearance, its peak expansion velocity, specific location
along the remnant's outer periphery, and the lack of any obvious counter-jet to the south.
Below, we briefly review some of the proposed theories for the jet's origin and  discuss 
them in relation to our dynamical measurement results.

\subsubsection{Plasma Instabilities Models } 

Several authors have suggested the jet is a post-SN feature resulting from an
interaction between the ejecta and the pulsar wind nebula. Many variations of
plasma instability models have been suggested including: 1) an aneurysm of
ejecta pushed aside by the pulsar's plasma flow \citep{byc75,ver85,mar90}, 2)
Rayleigh-Taylor instabilities \citep{che75,kun83} or differential acceleration
\citep{che75}, 3) pressure from swept-up magnetic field lines \citep{shu84}, 4)
ejecta entrainment in magnetic fields formed by highly-ordered outflows from
the polar caps of the pulsar itself \citep{ben84}, and 5) synchrotron driven
shocks \citep{san97}. 

A particularly attractive feature of such models is their ability to
explain both the non-radial appearance of the jet and the
lack of a counter-jet to the south. However, our proper motion measurements
indicate an essentially radial expanding jet. Moreover, 
some of these plasma instability models call for greater
acceleration of the jet's filaments than those found in the main body of the
nebula \citep{che75,shu84}, or a younger age compared to the rest of the nebula
\citep{kun83}. \textbf{ While a greater degree of acceleration or a younger age for the jet is 
not supported by our data which suggests a current jet age around 950 yr similar to the
Crab's historic age, a more modest expansion breakout acceleration is not completely  
ruled out by our proper motion measurements. }

\textbf{ For most plasma instability (`blowout') models, the exact location of a 
high-velocity ``jet'' along the remnant's outer boundary is
undefined but with a greater likelihood along  
the pulsar's NW--SE wind injection axis. }
However, the remnant's filamentary jet is some 45 degrees away from the pulsar's synchrotron jets,
which lie along a NW--SE axis (PA = 300 deg; \citealt{ng04}). Along this NW--SE
axis, the PWN appears to have pushed aside the remnant's outer ejecta filaments creating a
virtual hole in the filamentary ejecta along the NW limb \citep{law95,cad04}.
There the synchrotron emission can be seen to leak out of the
remnant's thick thermal plasma shell in optical, X-rays, and radio images
of the remnant. The lack of extened filamentary line emission ejecta in this
region similar to the northern `jet' is suggestive that the northern jet 
was not the result of a breakout of the pulsar's synchrotron wind nebula. 

\subsubsection{An Asymmetrical Explosion?}

\textbf{ Despite a lack of an obvious southern counter-jet, the remnant's northern jet might
represent high-velocity ejecta resulting from
a north-south bipolar expansion asymmetry.} In 1989, MacAlpine et al.
showed that N-S oriented spectral drift scans of the
Crab revealed a nearly N-S bipolar structure. The data showed that the pinched waist
of the  hourglass-like expansion  coincided with a band of helium
rich filaments (a ``high-helium band'') in which the computed helium mass fraction was
at least 75\% \citep{uom87}. \citet{mac89} suggested a helium rich circumstellar torus or disk 
could have resulted in constrained velocities in the horizontal region around
the progenitor. Subsequent drift scans of the remnant by \citet{fes97} and \citet{Smith03} 
confirmed the presence of a strong bipolar expansion.

Additional support for a circumstellar torus was presented by \citet{sch79} and
\citet{fes92} who examined the east and west indentations seen in 
the synchrotron emission (the synchrotron bays) and suggested these may have be caused by an E-W 
magnetic torus associated with the E-W band of He-rich filaments. Such a magnetic torus might  
constrain the east-west expansion of the PWN, 
leaving the bay indentations seen today.

In an asymmetrical explosion model for the jet, its location along the northern edge of
the remnant is not accidental \citep{fes93}.  The alignment of the jet
with the northern extent of the north expansion bubble seen in spectral drift 
scans of the remnant plus its major axis aligned nearly perpendicular
to the synchrotron bays suggest a causal link between the jet and
a possible constrained E-W expansion of the SN ejecta. 
The lack of an equally obvious southern `counter-jet' might be
explained by either a weaker ejection of material to the south and/or increased
ejecta confinement, and indeed drift scans of the remnant do show
higher velocities to the north compared to the south, 
relative to the pinched central zone \citep{mac89,fes97,Smith03}.  

The jet's filaments, being farther from the pulsar's energetic wind, would
experience less acceleration compared to knots and filaments located in the
main body of the nebula. Our measurements of other locations at the base of the
jet and to the east and west appear consistent with this picture, showing \textbf{larger} 
dynamical ages correlating with their increased distance from the pulsar (See
Table 3).

Finally, we note that our computed expansion age of the jet, although yielding an
expansion age consistent with the historic SN sighting in the year 1054, could
also be viewed consistent with \textbf{an initial} acceleration by the pulsar and \textbf{subsequent} 
deceleration through an interaction with a suspected outer
halo of unshocked ejecta \citep{che77,mur81,lun86,hes96}.  The presence of appreciable outer
ejecta, however, remains highly controversial with several radio, optical, and X-ray observations
failing to find any supportive evidence for its existence \citep{fra95,fes97,wal99,sew06}. 
\textbf{In any case, we view it as unlikely
that the degree of any PWN acceleration and possible outer ejecta induced deceleration 
would be so nearly offsetting as to lead to the currently derived jet expansion date 
of around C.E. 1054}. 

\section{Summary}

Proper motion analysis of 35 ejecta knots in the northern filamentary jet in the Crab
Nebula show the ejecta which form the jet to be less accelerated than the main
body of the nebula. Traced-back proper motions give an explosion date of 1055
$\pm$24 C.E. which lie in much closer agreement to the appearance of the 1054
C.E. historical guest star. The measure proper motions are largely radial
and directed away from the center of expansion. These data suggest the jet was
likely formed in the original explosion possibly as a result of bipolar SN kinematics
in which the E-W expansion of the ejecta was constrained by the presence of a
circumstellar disk. 

\section{Acknowledgments}

This research was partially funded by a DOF Research Grant, the ORL SS
program, and the Richter Memorial fund from Dartmouth College. 
We thank Molly Hammell for considerable programming assistance 
and Yuki Nakamura for reduction 
of the Suprime Cam image. This work is based in part on data collected at 
Subaru Telescope, which is operated by the National Astronomical
Observatory of Japan.

\clearpage

\end{document}